\documentclass{tMPH2e}

\newcommand*{\la}{\langle}
\newcommand*{\ra}{\rangle}

\usepackage{subfigure}

\begin{document}

\doi{10.1080/0026897YYxxxxxxxx}
 \issn{1362–3028}
\issnp{0026–8976}
\jvol{00}
\jnum{00} \jyear{2015} 

\articletype{Research Article}

\title{{\itshape Convergence properties of the multipole expansion 
of the exchange contribution to the interaction energy$\dagger$
\thanks{$^\dagger$ Dedicated to Professor Andreas Savin on the occasion 
of his 65 birthday.\vspace{3pt}}}}

\author{Piotr Gniewek$^{a}$$^{\ast}$\thanks{$^\ast$Corresponding author, 
e-mail: pgniewek@tiger.chem.uw.edu.pl
\vspace{6pt}} and Bogumi{\l} Jeziorski$^{a}$\\\vspace{6pt}  $^{a}${\em{Faculty of Chemistry, University of Warsaw, 
Pasteura 1, 02-093 Warsaw, Poland}}
\\\vspace{6pt}\received{\today } }

\maketitle

\begin{abstract}
The conventional surface integral formula $J_{\rm surf}[\Phi]$
and an alternative volume integral formula $J_{\rm var}[\Phi]$
are used to compute the asymptotic exchange splitting of the interaction
energy of the hydrogen atom and a proton employing the primitive function
$\Phi$ in the form of its truncated multipole expansion. Closed-form
formulas are obtained for the asymptotics of $J_{\rm surf}[\Phi_N]$
and $J_{\rm var}[\Phi_N]$, where $\Phi_N$ is the multipole expansion of $\Phi$
truncated after the $1/R^N$ term, $R$ being the internuclear separation.
It is shown that the obtained sequences of approximations converge
to the exact results with the rate corresponding to the convergence
radius equal to 2 and 4 when the surface and the volume integral formulas
are used, respectively. When the multipole expansion of a truncated, 
$K$th order polarization function is used to approximate the primitive 
function the convergence radius becomes equal to unity in the case 
of $J_{\textrm{var}}[\Phi]$. At low order the observed convergence 
of $J_{\rm var}[\Phi_N]$ is, however, geometric and switches to harmonic only 
at certain value of $N=N_c$ dependent on $K$. An equation for $N_c$ is derived 
which very well reproduces the observed $K$-dependent convergence pattern. 
The results shed new light on the convergence properties of the conventional 
SAPT expansion used in applications to many-electron diatomics.
\end{abstract}

\begin{figure}
\begin{center}
\begin{minipage}{13cm}
\vspace{2cm}
\resizebox*{13cm}{!}{\includegraphics{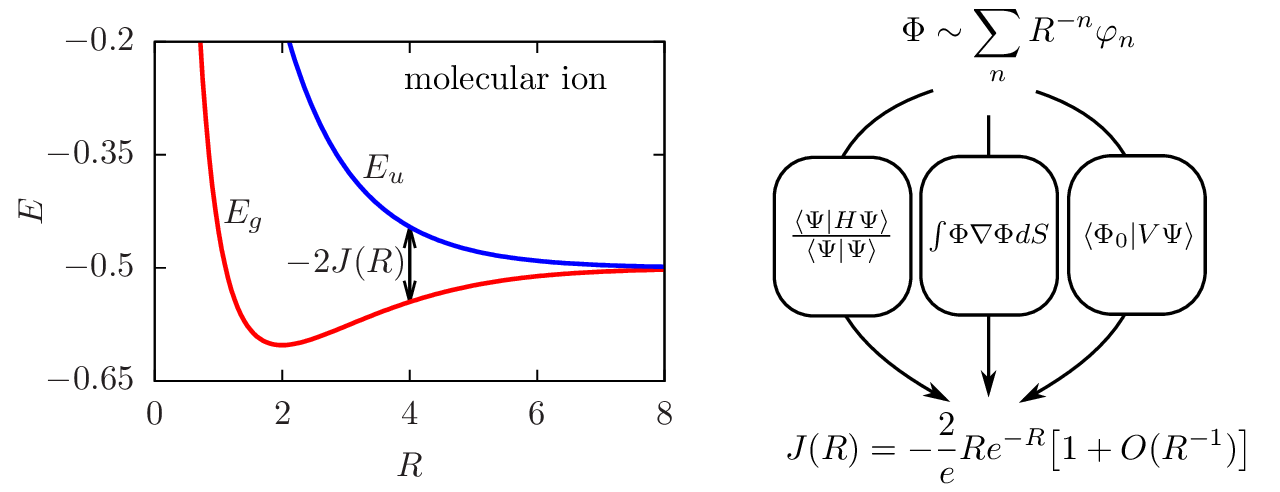}}%
\caption{Graphical abstract: The leading term of exchange energy can be calculated from the multipole expansion of the wave function
using variational principle, surface integral, or perturbation theory.  }%
\end{minipage}
\end{center}
\end{figure}

\section{Introduction}

Energetic effects of noncovalent molecular interactions are very small,
compared to the energies of noninteracting subsystems, nevertheless
they determine the properties of condensed matter \cite{Kaplan:06,Stone:13} 
and influence chemical reactivity \cite{Kim:15}.
The theoretical framework for the understanding of the nature and origin 
of noncovalent interactions is provided by Symmetry Adapted Perturbation Theory
(SAPT) \cite{Jeziorski:94,Szalewicz:05}, which introduces rigorous decomposition 
of the interaction energy into electrostatic, induction, dispersion and exchange 
contributions. 
At large interatomic distances $R$ the first three of these constituents can be
expanded in powers of $R^{-1}$, leading to the so-called multipole expansion of
interaction energy 
\begin{equation}\label{eq:Eint_asymptot_exp}
    E_{\textrm{int}}(R) \sim \sum_{n} C_n R^{-n} ,
\end{equation}
with  the $C_n$'s coefficients referred to as the van der Waals constants. 
While the expansion of Eq.~(\ref{eq:Eint_asymptot_exp}) is divergent for any $R$ 
\cite{Young:75,Cizek:86}, 
it is known to provide the correct asymptotic representation of $E_{\textrm{int}}(R)$
for large $R$ \cite{Ahlrichs:76,Morgan:80}. 
From the perspective of the large $R$ asymptotics of $E_{\textrm{int}}$, 
the exchange effects might seem to be insignificant. However, at intermediate 
and short distances the importance of the exchange part of $E_{\textrm{int}}$ 
is hard to overestimate, as it is responsible for the splitting 
of potential energy curves and for the strong repulsion in the case 
of interaction of closed-shell atoms when it    
provides the necessary quenching of the attractive induction and dispersion 
components of the interaction energy.

The exchange effects are caused by the resonant tunneling of one or more 
electrons between the interacting systems and are considered to be non-perturbative,
that is, an application of the straightforward Rayleigh-Schr\"odinger 
perturbation treatment is unable to provide even qualitative description 
of the exchange contribution to $E_{\textrm{int}}$ 
\cite{Chalasinski:77,Cwiok:92:Pol}. 
This problem can be circumvented by the symmetry forcing procedure \cite{Jeziorski:77},
in which the wave function is correctly symmetrized in each order of perturbation 
theory. Different variants of symmetry forcing lead to different 
SAPT expansions \cite{Jeziorski:77, Kutzelnigg:78}, of which the most important is the 
Symmetrized Rayleigh-Schr\"odinger (SRS) perturbation approach  
\cite{Chalasinski:77,Jeziorski:78}. 
SRS is nowadays routinely applied in studies of noncovalent interactions of many-electron systems
\cite{Szalewicz:12,Sherrill:12,Jansen:14,Hesselmann:14}. 

The success of SRS approach is somewhat bewildering, as it is based on the so-called 
polarization approximation (PA), which is the straightforward application 
of the Rayleigh-Schr\"odinger 
(RS) perturbation theory to molecular interactions.
PA is known to diverge when at least one of 
the interacting subsystems has more than two electrons \cite{Kutzelnigg:80,Adams:96,Patkowski:01,Przybytek:04}. 
On the other hand, PA converges for H$_2^+$ and H$_2$ to the ground state wave function 
and energy \cite{Chalasinski:77, Cwiok:92:Pol, Kutzelnigg:92}. 
For large $R$ this convergence is very slow, as the convergence radius of PA 
exceeds one by a quantity exponentially decaying with $R$ 
\cite{JeziorskiSchwalmSzalewicz1980, Cwiok:92:Pol, Kutzelnigg:92, Gniewek:14}. 
Irrespectively of the ultimate convergence or divergence, 
PA provides the asymptotic approximation to the primitive
function \cite{Kutzelnigg:80} $\Phi$ of the interacting system, in the sense that
\cite{Ahlrichs:76,Jeziorski:82}
\begin{equation}\label{eq:Phi_K_as_asympt_exp}
    \Psi_\nu = \mathcal{A}_\nu \Phi^{(K)}+ O(R^{-\kappa(K+1)}) ,
\end{equation}
where $\Psi_\nu$ is the wave function of the state $\nu$, $\mathcal{A}_\nu$ 
is the symmetry 
operator associated with this state, $\Phi^{(K)}$ is the PA wave function 
computed through the $K$th order, and $\kappa$ equals two if at least one 
of the interacting subsystems is charged and three if both are neutral.

From the practical point of view it is important to know the limits of validity and 
convergence properties of various approaches to calculation of the exchange contribution 
to $E_{\textrm{int}}$. 
In this paper we present exact results concerning the rate 
of convergence of the exchange energy $J(R)$ of the hydrogen 
molecular ion H$^+_2$ when computed using the multipole expansion of the 
primitive wave function.  
The exchange energy is defined in this context as 
\begin{equation}\label{eq:J_definition}
    J(R) = \tfrac{1}{2} \big [ E_g(R) - E_u(R) \big ] ,
\end{equation}
where $E_g$ and $E_u$ are the Born-Oppenheimer energies of the lowest \emph{gerade} and
\emph{ungerade} states, respectively.
The H$_2^+$ ion is particularly suitable for studies of the convergence of exchange effects 
for two reasons: ($a$) the analysis is simplified by the fact that there are no electron correlation effects involved and ($b$) 
the asymptotic expansion of $J(R)$ is known exactly for this 
system \cite{Cizek:86}.  
H$_2^+$ is one of the most important models of the theory of molecular interactions 
and there is a wealth of analytical results available for this species 
\cite{Chipman:73, Whitton:76, Kutzelnigg:78, Damburg:84, Cizek:86}. 
H$_2^+$ is also a prototypical double-well system \cite{Harrell:80, Damburg:82}. 
The asymptotic expansion of exchange energy of H$_2^+$ can be conveniently written in the form 
\begin{equation}\label{eq:J_asymptotics}
    J(R) = \frac{2}{e}\,R e^{-R}( j_0  +  j_1\, R^{-1}  +  j_2 \,R^{-2}  +  \dots ) ,
\end{equation}
where $j_0 = -1, j_1 = -\frac{1}{2}$, and $ j_2 = -\frac{25}{8}$. The higher  
$j_n$ coefficients can be found in Ref. \cite{Cizek:86}.  

The most widely used method of calculating $J(R)$ employs the surface-integral formula 
$J_{\textrm{surf}}[\Phi]$ given by 
\cite{Firsov:51,Holstein:52,Herring:62,Tang:98}
\begin{equation}\label{eq:J_surf_def}
    J_{ \textrm{surf} } [ \Phi ] = \frac{ \int_M \Phi \nabla \Phi d \mathbf{S} }{
    \la \Phi | \Phi \ra  - 2 \int_{ \textrm{right} } \Phi^2 d V },
\end{equation}
where the primitive function $\Phi$ is assumed to be localized in the ``left'' half of space
and $M$ is the median plane dividing the space into ``left'' and ``right'' halves.
Atomic units ($\hbar=e=m_e=1$) are used in Eq.~(\ref{eq:J_surf_def}) and
throughout the paper.

Alternatively, the exchange energy can be calculated from the SAPT formula 
\cite{Gniewek:14} 
\begin{equation}\label{eq:J_SAPT_def}
    J_{ \textrm{SAPT} } [ \Phi ] = \frac{  
    \la \varphi_0 | V P  \Phi \ra  \la \varphi_0 | \Phi \ra - \la \varphi_0 | V \Phi \ra \la \varphi_0 | P  \Phi \ra 
    }{ \la \varphi_0 | \Phi \ra^2 - \la \varphi_0 | P \Phi \ra^2 } ,
\end{equation}
where $V$ is the interaction operator, $\varphi_0$ is the unperturbed 
wave function, and $P$ is the operator inverting the electronic coordinates 
with respect to the midpoint of the internuclear axis.
When $\Phi$ is approximated by the polarization function $\Phi^{(K)}$ 
and the r.h.s. of Eq. (\ref{eq:J_SAPT_def}) is expanded in powers of $V$, 
one obtains the SRS expansion 
of the exchange energy through the $K$th order.

In our recent work \cite{Gniewek:15} we showed that 
the exchange energy can be very accurately 
calculated from the variational principle based formula $J_{\textrm{var}}[\Phi]$
\begin{equation}\label{eq:J_var_def}
    J_{ \textrm{var} } [ \Phi ] = \frac{ 
    \la \Phi | H P  \Phi \ra \la \Phi | \Phi \ra - \la \Phi | H \Phi \ra \la \Phi | P  \Phi \ra 
    }{   \la \Phi | \Phi \ra^2 - \la \Phi | P \Phi \ra^2 },
\end{equation}
where $H$ is the full Hamiltonian of the interacting system.
Numerical results illustrating the convergence of $J_{\textrm{surf}}[\Phi]$, 
$J_{\textrm{SAPT}}[\Phi]$, and $J_{\textrm{var}}[\Phi]$ 
with respect to the order of multipole expansion used to approximate 
the primitive wave function $\Phi$ were 
presented in our previous work \cite{Gniewek:15}.
By analyzing these numerical results we have found that 
the convergence rate of the leading term, $j_0$, 
of the asymptotic expansion of Eq. (\ref{eq:J_asymptotics}) corresponds 
to the convergence radius equal to 2, 1, and 4 for the surface-integral, 
SAPT, and variational formula, respectively.

In this communication we shall present a rigorous mathematical derivation 
of the discovered convergence rates and the corresponding  
convergence radii (with respect to 
the order of multipole expansion of $\Phi$) for $j_0$ calculated from 
the surface $J_{\textrm{surf}}[\Phi]$ and variational $J_{\textrm{var}}[\Phi]$ 
expression (in the case of $J_{\textrm{SAPT}}[\Phi]$  
such analysis has already been presented  in Ref.~\cite{Gniewek:15}). 
Furthermore we derive closed-form formulas for partial sums 
of the series approximating  $j_0$ and show that these partial sums 
converge to the correct value $j_0=-1$.
We shall also explain why in the calculations of $j_0$ using 
the primitive function approximated by
 a finite polarization expansion $\Phi^{(K)}$ the convergence 
is initially geometric but switches to harmonic when the multipole 
expansion of $\Phi^{(K)}$ 
is carried out to sufficiently high order.

\section{Asymptotic expansion of the polarization wave function}

The non-relativistic Born-Oppenheimer Hamiltonian for an 
interacting system consisting of two molecules A and B decomposes 
naturally as $H = H_0 + V$, where $H_0 = H_A + H_B$ is 
the unperturbed part 
($H_A$ and $H_B$ being the Hamiltonians of isolated monomers A and B, 
respectively) and $V$ is the perturbation accounting for the Coulomb 
interactions between charged particles belonging to different monomers. 
In the case of H$_2^+$ the pertinent partitioning of $H$ is
\begin{equation}\label{eq:H0_and_V}
    H_0 = -\frac{1}{2} \nabla^2 - \frac{1}{r_a}, 
    \quad\quad V = \frac{1}{R} -\frac{1}{r_b} , 
\end{equation}
where $r_a$ and $r_b$ are the distances of the electron from the 
nuclei $a$ and $b$, respectively.

Polarization approximation is obtained when one applies the 
usual RS perturbation 
theory to the above definition of $H_0$ and $V$, with the 
the unperturbed wave function $\varphi_0$ taken as 
the product of the ground state wave functions of isolated 
subsystems (in case of H$_2^+$ the unperturbed wave function is 
the ground state wave function of a hydrogen atom $a$, $\varphi_0 = 1s_a$). 
The polarization corrections to energy are given by
\begin{equation}\label{eq:E_RS_corrections}
    E^{(k)} = \la \varphi_0 | V \varphi^{(k-1)} \ra ,
\end{equation}
whereas the wave function corrections are defined recursively 
by equations
\begin{equation}\label{eq:phi_RS_corrections}
    ( H_0 - E_0 ) \varphi^{(k)} = - V \varphi^{(k-1)} 
    + \sum_{m=1}^k E^{(m)} \varphi^{(k-m)} 
\end{equation}
and the intermediate normalization condition 
$\la \varphi^{(k)} | \varphi_0 \ra = 0$ holding for $k>0$. 
For $k=0$ we have $\varphi^{(0)} = \varphi_0$.
The wave function corrections can be written in a product form, 
separating the zeroth-order wave function, 
$\varphi^{(k)} = \varphi_0 f^{(k)}$, 
where $f^{(0)}=1$ and for $k\ge 1$ the factors $f^{(k)}$ satisfy the equation 
 \cite{Dalgarno:55,Tang:91}
\begin{equation}\label{eq:fn_recurrence_full}
    -\tfrac{1}{2} \nabla^2 f^{(k)} + \frac{ \partial f^{(k)} }{ \partial r_a } 
    = - V f^{(k-1)} + \sum_{m=1}^k E^{(m)} f^{(k-m)} .
\end{equation}
When the perturbation $V$ in Eq. (\ref{eq:fn_recurrence_full}) is represented 
by its multipole expansion 
\begin{align}\label{eq:V_multipole_exp}
   V \sim \, \,  & V_2 \, R^{-2}   + V_3 \, R^{-3}  + V_4 \, R^{-4}  +   \ldots , \\
           &  V_n  =  - r_a^{n-1} P_{n-1}(\cos \theta_a) ,
\end{align}
one obtains the asymptotic expansion for $f^{(k)}$ 
\begin{equation}\label{eq:fn_expansion_1}
   f^{(k)} \sim \sum_{n} f^{(k)}_{n}(r_a, \theta_a)\, R^{-n} ,
\end{equation}
where $\theta_a$ is the inclination angle of the electron 
in a coordinate system centered at the nucleus $a$ and $P_l(x)$ 
is the  Legendre polynomial. 
Using Eq. (\ref{eq:fn_recurrence_full}) one can verify that   
the factors  $f^{(k)}_{n}(r_a, \theta_a)$ are finite order polynomials 
in $r_a$ and $\cos\theta_a$
\begin{equation}\label{eq:fn_expansion_2}
 f^{(k)}_{n}(r_a, \theta_a)= 
\sum_{m}  r_a^m  \sum_{l=0} t^{(k)}_{nml}\,  P_l(\cos\theta_a) .
\end{equation}
such that $m \le n$ and $l \le m$.  Eq. (\ref{eq:fn_expansion_2}) 
may be viewed as an expansion of Eq. (28) of Ref.~\cite{Gniewek:15}
in powers of the perturbation $V$ [the function $f_n$ in Eq. (28) 
of Ref.~\cite{Gniewek:15} is defined via $\phi_n= f_n\phi_0$; this 
definition has been erroneously omitted in the line preceding 
Eq. (28)].

Since $\la \Phi | \Phi \ra = 1 + O(R^{-4})$ and 
$\la \Phi | H \Phi \ra = E_0 + O(R^{-4})$ the asymptotically leading term of 
$J_{\textrm{var}}[\Phi]$, 
defined by the $j_0$ coefficient in Eq.~(\ref{eq:J_asymptotics}),
can be obtained by considering a  simplified formula
\begin{equation}\label{eq:J_var_star}
    J_{\textrm{var}}^*[\Phi] = \la \Phi | ( H - E_0 ) P \Phi \ra .
\end{equation}
Evaluation of Eq.~(\ref{eq:J_var_star}) with the multipole expansion 
of the polarization function  truncated after the $K$th term,
\begin{equation}\label{PhiK}
    \Phi^{(K)} \sim \varphi_{0} \sum_n R^{-n} \sum_{k=0}^K f^{(k)}_{n} ,
\end{equation}
leads to a linear combination of integrals 
\begin{equation}\label{eq:master_int_asymptot}\begin{aligned}
    & \frac{1}{\pi}\int e^{-r_a-r_b} r_a^m r_b^{m'} P_l(\cos\theta_a) P_{l'}(\cos\theta_b) d^3{\bf  r} = \\
    & \quad = 2 e^{-R} \frac{ (m+1)! (m'+1)! }{ (m+m'+3)! } R^{m+m'+2} \big [ 1 + O( R^{-1} ) \big ] .
\end{aligned}\end{equation}
The asymptotic formula (\ref{eq:master_int_asymptot}) 
shows that only the terms with $m=n$ in Eq.~(\ref{eq:fn_expansion_2})
contribute to $j_0$ of the expansion~(\ref{eq:J_asymptotics}).
For this reason we will further consider only the dominant, $m=n$ part 
of $f^{(k)}_n$, denoted $\widetilde f^{(k)}_n$, given by  
\begin{equation}\label{eq:fn_expansion_simplified}
    \widetilde f^{(k)}_n = r_a^n \sum_{l=0}t_{nl}^{(k)} \, 
    P_l(\cos\theta_a) ,
\end{equation}
where $t_{nnl}^{(k)}$ has been denoted by $t_{nl}^{(k)}$.  
Inserting Eqs. (\ref{eq:fn_expansion_1}) and (\ref{eq:fn_expansion_2})
into Eq.~(\ref{eq:fn_recurrence_full}) and comparing coefficients 
at terms proportional to $r_a^{n-1} R^{-n}$ one obtains the relation
\begin{equation}\label{eq:t_k_n_recurrence_Legendre_P}
    n \sum_{l=0}^{n-1} t^{(k)}_{nl} P_l(\cos\theta_a) 
    = \sum_{j=0}^{n-2} \sum_{l=0} t^{(k-1)}_{jl} P_l (\cos\theta_a)
 P_{n-j-1}(\cos\theta_a) .
\end{equation}
In deriving Eq. (\ref{eq:t_k_n_recurrence_Legendre_P}) we took advantage 
of the fact that $\nabla^2 f^{(k)}$ and the last term in 
Eq. (\ref{eq:fn_recurrence_full}) do not contain terms  
proportional to $r_a^{n-1} R^{-n}$. 

Since the asymptotic value of the integral 
of Eq.~(\ref{eq:master_int_asymptot}) is independent of $l$ and $l'$, 
the  angular dependence of $\widetilde f^{(k)}$ does not affect 
the asymptotic estimate of Eq. (\ref{eq:J_var_star}) and, therefore, 
has no influence on $j_0$.
In other words, the value of the asymptotically leading term of 
$J_{\textrm{var}}[\Phi]$ depends only on the values of $\Phi$
on the line joining the nuclei $a$ and $b$, where 
$P_l(\cos\theta_a)= P_{l'}(\cos\theta_b) =1$, 
similarly as it was 
observed for $J_{\textrm{SAPT}}[\Phi]$ in Ref.~\cite{Gniewek:15}.
Thus, in the analysis of the convergence properties 
of approximations to $j_0$ it is sufficient to calculate 
the values of the functions $ \widetilde f^{(k)}_n$ 
on the line joining the nuclei,
\begin{equation}
    \widetilde f^{(k)}_n \big |_{\theta_a=0} = t_{n}^{(k)} r_a^n , 
\end{equation}
where $ t^{(k)}_n $ is defined as the sum 
\begin{equation}
    t^{(k)}_n = \sum_{l=0} t^{(k)}_{nl} .
\end{equation}
Substituting $\theta_a = 0$ in Eq.~(\ref{eq:t_k_n_recurrence_Legendre_P}) 
we obtain a simple recurrence relation for the $t^{(k)}_n$ coefficients 
\begin{equation}\label{eq:t_k_n_recurrence}
    t_{n}^{(k)} = \frac{1}{n} \sum_{j=2k-2}^{n-2} t_{j}^{(k-1)} 
\end{equation}
with the initial, $k=0$ values given by  $t^{(0)}_0 = 1$ and
$t^{(0)}_n = 0$ for $n > 0$.
In setting the lower summation limit in 
Eq. (\ref{eq:t_k_n_recurrence}) we used the fact that 
$t^{(k)}_n = 0$ for $n < 2k$.
In Eq.~(\ref{eq:t_k_n_recurrence}) and throughout the paper we use the convention that when 
the lower summation limit is greater than the upper one the sum is zero. 
Using  Eq.~(\ref{eq:t_k_n_recurrence}) it is straightforward to show that 
\begin{equation}\label{eq:t_k_n_recurrence_fast}
    t_n^{(k)} = \frac{ n-1 }{ n } t_{n-1}^{(k)} + \frac{1}{n} t_{n-2}^{(k-1)} , \quad \quad n > 1 ,
\end{equation}
which furnishes a fast algorithm for the calculation of $t^{(k)}_n$. 

The values of the multipole expansion of $\Phi^{(K)}$, 
Eq. (\ref{PhiK}),  truncated after $R^{-N}$ 
term and computed on the line joining the nuclei with  $f^{(k)}_n$
replaced by its dominant part $\widetilde f^{(k)}_n$,
Eq. (\ref{eq:fn_expansion_simplified}),  are given by
\begin{equation}\label{eq:Phi_K_N_expansion}
 \widetilde   \Phi^{(K)}_N \big |_{\theta_a=0} 
= \varphi_0 \sum_{n=0}^N d_n^{(K)} r_a^n \,  R^{-n} ,
\end{equation}
where
\begin{equation}\label{eq:d_K_N_definition}
    d_n^{(K)} = \sum_{k=0}^K t_n^{(k)} .
\end{equation}
The multipole expansion of the primitive function $\Phi$ truncated after  
$R^{-N}$, denoted by $\Phi_N$,  is 
\begin{equation}\label{eq:Phi_N_def}
 \Phi_N = \lim_{ K \rightarrow \infty } \Phi^{(K)}_N 
     = \varphi_0 \sum_{n=0}^N R^{-n} \sum_{k=0}^{[n/2]} f^{(k)}_n , 
\end{equation}
where $[n/2]$ denotes the integral part of $n/2$. The finite 
upper limit of the 
$k$ summation in Eq. (\ref{eq:Phi_N_def}) is a consequence of the fact 
that $t_n^{(k)} = 0$ 
for $k > n/2$. The dominant part  of $\Phi_N$, denoted by  $\widetilde \Phi_N$,
is defined in the same way as $\Phi_N$ except that $f^{(k)}_n$ is 
replaced by $\widetilde f^{(k)}_n$. 
On the line joining the nuclei the function $\widetilde \Phi_N$ 
is given by Eq.~(\ref{eq:Phi_K_N_expansion})
with superscripts $(K)$ omitted and $d_n$ defined by 
\begin{equation}\label{eq:d_N_short_sum}
    d_n = \lim_{ K \rightarrow \infty } d^{(K)}_n =
\sum_{k=0}^{[n/2]} t_n^{(k)} . 
\end{equation}

We shall now derive a closed-form formula for $d_n$. 
This can be done via a recurrence for differences of consecutive $d_n$'s
with $n \ge 2$, which using Eqs.~(\ref{eq:t_k_n_recurrence})
and~(\ref{eq:d_N_short_sum}) takes the form  
\begin{equation}\label{eq:dN_dNm1_2}
    d_n - d_{n-1} = \sum_{k=1}^{\infty} 
 \bigg ( \frac{1}{n} \sum_{j=2k-2}^{n-2} t_{j}^{(k-1)}
 - \frac{1}{n-1} \sum_{j=2k-2}^{n-3} t_{j}^{(k-1)} \bigg ) ,
\end{equation}
where the lower limit of the $k$ summation was set equal to 
1 since  for $n \ge 1$ the $k=0$ term does not contribute 
to the summation in Eq. (\ref{eq:d_N_short_sum}).
The summation over $k$ is actually finite and limited by the condition 
$t_n^{(k)} = 0$ for $k > n/2$. 

Extracting the $j=n-2$ term form the left inner sum and combining the 
remaining part with the other sum gives 
\begin{equation}\label{eq:dnN_dNm1_3}
    d_n - d_{n-1} = \frac{1}{n} \sum_{k=1}^{\infty} 
\bigg ( t_{n-2}^{(k-1)} - 
\frac{1}{ n-1 } \sum_{j=2k-2}^{n-3} t_{j}^{(k-1)} \bigg ) .
\end{equation}
Comparison with Eqs.~(\ref{eq:t_k_n_recurrence})~and~(\ref{eq:d_K_N_definition}) allows to rewrite Eq.~(\ref{eq:dnN_dNm1_3}) as
\begin{equation}
    d_n - d_{n-1} = - \frac{ 1 }{ n } ( d_{n-1} - d_{n-2} ) ,
\end{equation}
which together with the fact that $d_0 = 1$ and $d_1 = 0$ gives 
\begin{equation}\label{eq:d_N_formula}
    d_n = \sum_{m=0}^n \frac{ (-1)^m }{ m! } .
\end{equation}
Eq. (\ref{eq:d_N_formula}) agrees with Eq. (36) of Ref.~\cite{Gniewek:15} 
obtained using a different derivation, not related to the polarization
expansion of the wave function.

One may note that after multiplication by $n!$ 
the coefficients $d_n$ and $t_n^{(k)}$ 
have a combinatorial interpretation: $n!\,d_n$ is equal to
the so-called derangement number $D_n$ (see Ref.~\cite{Riordan:14}~p.~59, 
Ref.~\cite{Comtet:74}~p.~180, and Ref.~\cite{OEIS} entry A000166) 
defined as the number of permutations with no fixed points 
(ie. permutations in which all elements change places), while  
$n!\, t_n^{(k)}$ is equal to the associated Stirling number $d(n,k)$, 
defined as  the number of derengements of $n$ elements having exactly 
$k$ permutation cycles (Ref.~\cite{Comtet:74}~p.~256, 
Ref.~\cite{Riordan:14}~p.~73, and Ref.~\cite{OEIS} entry A008306).

\section{Closed-from expressions for the partial sums 
          and the convergence rates 
         of the multipole expansion of the exchange energy $j_0$}

Approximations to  $j_0$ obtained from $J_{\textrm{surf}}[\Psi]$ 
and $J_{\textrm{var}}[\Psi]$, where $\Psi$ is any approximation  
to the primitive function $\Phi$,  will be denoted by 
$j_0^{\textrm{surf}}[\Psi]$ and $j_0^{\textrm{var}}[\Psi]$, respectively. 
The convergence of these approximations resulting when $\Psi$ 
is substituted by the multipole expanded polarization function 
can be characterized with the help of the increment ratios 
defined as
\begin{equation}\label{eq:rho_K_N_j0surf}
    \rho_N^{(K)}(j^{\textrm{surf}}_0) 
    =  \frac{   j_0^{\textrm{surf}}[ \Phi^{(K)}_{N}] - j_0^{\textrm{surf}}[ \Phi^{(K)}_{N-1} ] 
    }{   j_0^{\textrm{surf}}[\Phi^{(K)}_{N+1}]-j_0^{\textrm{surf}}[\Phi^{(K)}_N]   } ,
\end{equation}
where $\Phi^{(K)}_N$ is the multipole expansion of $\Phi^{(K)}$ truncated
after the $R^{-N}$ term, cf. Eq. (\ref{PhiK}).  
Increment ratios $\rho_N^{(K)}(j^{\textrm{var}}_0)$ are defined analogously. 
The convergence of results obtained when $\Psi$ is substituted by  $\Phi_N$ 
will be characterized by the increment rations  
$\rho_N(j^{\textrm{surf}}_0)$ and $\rho_N(j^{\textrm{var}}_0)$, defined 
as in Eq.~(\ref{eq:rho_K_N_j0surf}), but with superscripts 
$(K)$ omitted. 
Similar convention will be applied to other quantities as well --- 
symbols without the superscript $(K)$ will denote the 
$K \rightarrow \infty$ limits 
of the corresponding expressions carrying this superscript.
It should be noted that the r.h.s. of Eq.~(\ref{eq:rho_K_N_j0surf}) 
remains unchanged if the functions $\Phi^{(K)}_N$ or $\Phi_N$ are 
replaced by $\widetilde{\Phi}^{(K)}_N$
or  $\widetilde{\Phi}_N$, respectively.

\subsection{Surface-integral formula}

It has been shown in  Ref.~\cite{Tang:91} 
that the large $R$ asymptotics of $J_{\rm surf}[\Phi]$
is determined by the value of $\Phi$ at the mid-point 
of the line joining the nuclei. Specifically, if 
$\Phi=\phi_0\, f(r_a,\theta_a)$
then 
\begin{equation} 
J_{\rm surf}[\Phi] = - \frac{1}{2} R e^{-R} \, [f(R/2,0)]^2 + O(e^{-R}).
\end{equation} 
Thus, in view of Eq. (\ref{eq:Phi_K_N_expansion}) 
the asymptotics of  $J_{\rm surf}[\Phi_N^{(K)}]$ is determined 
by the constant  
\begin{equation}\label{eq:j0surf_PhiN_via_rN}
     j_0^{\textrm{surf}} [ \Phi_N^{(K)}]= 
     j_0^{\textrm{surf}} [ \widetilde{\Phi}_N^{(K)} 
] = -\frac{e}{4} \, \big( r_N^{(K)} \big )^2 ,
\end{equation}
where 
\begin{equation}\label{eq:rNK_def}
    r_N^{(K)} = \sum_{n=0}^N \frac{ d^{(K)}_n }{ 2^n } .
\end{equation}
Using mathematical induction one can easily prove that the 
$K\rightarrow\infty$ limit of $ r_N^{(K)}$, denoted by $ r_N$,  is given by 
\begin{equation}\label{eq:r_N_closed_form}
    r_N = 2 \, e_N \big ( -\tfrac{1}{2} \big ) - 2^{-N} d_N , 
\end{equation}
where $e_n(x)$ denotes the exponential sum function
\begin{equation}
    e_n(x) = \sum_{m=0}^n \frac{ x^m }{ m! } . 
\end{equation}
Combining  Eqs.~(\ref{eq:j0surf_PhiN_via_rN}) 
and~(\ref{eq:r_N_closed_form}) one obtains a simple closed-form formula 
for $j_0^{\textrm{surf}}[\Phi_N]$, that is  
for the exchange energy determined by partial sums 
of the multipole expansion of the primitive function: 
\begin{equation}\label{eq:j0surf_Nth_sum}
     j_0^{\textrm{surf}}[\Phi_N] 
    = -\frac{e}{4} \, \big [ 2 e_N(-\tfrac{1}{2}) - 2^{-N} d_N \big ]^2 .
\end{equation}
Taking the $N \rightarrow \infty$ limit we see that the leading, $j_0$ term
in the expansion of  Eq. (\ref{eq:J_asymptotics}) is correctly 
obtained in this way, i.e,
\begin{equation}
    \lim_{N \rightarrow \infty} j_0^{\textrm{surf}}[\Phi_N] = -1 , 
\end{equation}
in agreement with the results of Ref.~\cite{Gniewek:15}. 

Equation (\ref{eq:j0surf_Nth_sum}) can be compared with the
formula for $j_0$ obtained by Tang et al. \cite{Tang:91} 
using the surface-integral expression  
and the polarization function $\Phi^{(K)}$. 
Their Eq.~(7.25) can be recast in our notation as follows:
\begin{equation}\label{eq:Tang:91_7.25}\begin{aligned}
    j_0^{\textrm{surf}}[\Phi^{(K)}] & =
 -\frac{1}{2} \bigg [  1  
    +  \sum_{m_1 = 2}^\infty \frac{ 1 }{ m_1 2^{m_1}}
  +  \sum_{m_1,m_2 = 2}^\infty \frac{ 1 }{ m_1 ( m_1 + m_2 ) 2^{m_1+m_2} }  \\
    & +  \sum_{m_1,m_2,m_3 = 2}^\infty \frac{ 1 }{ m_1 ( m_1 + m_2 )( m_1 + m_2 +m_3)
 2^{m_1+m_2+m_3} } 
    + \dots \bigg ]^2 ,
\end{aligned}\end{equation}
where the terms in the square brackets  
involving $k$tuple summation originate from the $k$th order polarization function 
$\phi^{(k)}$. The last term (not explicitly written) involves the $K$tuple summation and 
originates from $\phi^{(K)}$. 
Eq.~(\ref{eq:Tang:91_7.25}) is a sum of terms of increasingly high 
order in $V$, each of which is an infinite sum corresponding to the expansion in powers 
of $R^{-1}$. 
On the other hand, Eq.~(\ref{eq:j0surf_Nth_sum}) contains only single sums over 
terms of different order in $R^{-1}$, each of which is a closed-form sum of contributions of 
different orders in $V$.

Knowing the expression for $r_N$, 
the increment ratio $\rho_N( j^{\textrm{surf}}_0 )$, 
defined by the $K\rightarrow\infty$ limit of Eq.~(\ref{eq:rho_K_N_j0surf}), 
can be written in the form 
\begin{equation}\label{eq:rho_j0_surf_K_infty}
    \rho_N( j^{\textrm{surf}}_0 ) = 
    2\, \frac{ d_N}{ d_{N+1}} \,
      \frac{ r_N - 2^{-N-1} d_N }{ r_N + 2^{-N-2} d_{N+1} } .
\end{equation}
When $N \rightarrow \infty$ the ratio $ d_N/ d_{N+1}$ is equal 
to 1 with an error of the order of $1/(N+1)!$
Estimating the remaining factor using Eq.~(\ref{eq:r_N_closed_form})
one finds that 
\begin{equation}\label{eq:rho_N_j0surf}
    \rho_N ( j_0^{\textrm{surf}} ) = 2 - \frac{ 3 }{ 4 \sqrt{e} } 2^{-N} + O(4^{-N}) ,
\end{equation}
which is the result discovered numerically in Ref.~\cite{Gniewek:15}.
We see that the sequence of approximations to 
$j_0$ obtained using the surface integral
formula and the multipole expansion of the wave function converges 
like a series with the convergence radius equal to 2. In Sec. 4.1 
we show that for finite $K$ the convergence radius corresponding to 
the sequence $ \rho_N^{(K}) ( j_0^{\textrm{surf}} )$ remains also
equal to 2, although the rate of convergence becomes somewhat slower 
in this case.  

Concluding this Subsection it may be remarked that the authors 
of Ref.~\cite{Tang:91} were only concerned with the 
$K\rightarrow \infty$ limit of Eq. (\ref{eq:Tang:91_7.25})
and did not consider the convergence rate of the obtained series
expansions. 
In fact, this rate cannot be simply inferred 
from Eq.~(\ref{eq:Tang:91_7.25}).
On the other hand, the convergence rate corresponding 
to Eq.~(\ref{eq:j0surf_Nth_sum}) can be easily obtained and 
is given by Eq.~(\ref{eq:rho_N_j0surf}).

\subsection{Variational formula}

In the case of the variational volume-integral formula it is easier
to analyse the convergence rate than to obtain a closed-form expression 
for the partial sum of the corresponding expansion. We therefore start by 
considering the increment ratio of Eq. (\ref{eq:rho_K_N_j0surf}) first.

\subsubsection{Convergence rate of the variational formula}  

Employing Eqs. (\ref{eq:master_int_asymptot}) and
 (\ref{eq:Phi_K_N_expansion}) one obtains 
\begin{equation}
    \frac{2}{e} j_0^{\textrm{var}} [ \Phi_N^{(K)} ] 
    = \sum_{m_1,m_2=0}^N d^{(K)}_{m_1} d^{(K)}_{m_2} G_{m_1,m_2} ,
\end{equation}
where 
\begin{equation}
    G_{m_1,m_2} = 2 \frac{ m_1! m_2! ( m_1 m_2 [m_1+m_2+4] - 2 ) }{ (m_1+m_2+3)! } .
\end{equation}
The numerator in  Eq. (\ref{eq:rho_K_N_j0surf}) multiplied by $2/e$ 
can now be written  as
\begin{equation}
    q^{(K)}_N = 2 d_N^{(K)} p_N^{(K)} + \big ( d_N^{(K)} \big )^2 G_{N,N} ,
\end{equation}
where 
\begin{equation}\label{eq:p_N_K_def}
    p_N^{(K)} = \sum_{m=0}^{N-1} d_m^{(K)} G_{m,N} .
\end{equation}

A closed-form formula for the the 
$K \rightarrow \infty$ limits of $p^{(K)}_N$, denoted by 
$p_N$,  can be obtained using the summation 
by parts formula, 
\begin{equation}\label{eq:summation_by_parts}
    \sum_{j=0}^n u_j w_j = u_n W_n 
    - \sum_{j=0}^{n-1} \big ( u_{j+1} - u_j \big ) W_j ,
\end{equation}
where
\begin{equation}
    W_j = \sum_{m=0}^j w_m .
\end{equation}
We shall use this formula with  $u_j = d_j$ and  $w_j = G_{j,N}$. In this case  
$    u_{j+1} - u_j = (-1)^{j+1}/(j+1)!$ and 
\begin{equation}\label{eq:Wj_summation_of_p}
    W_j = \frac{ 2 }{ (N+1)(N+2) } 
    - 2 \big [ (j+1)(N+5) + j^2 \big ] \frac{ (j+1)! N! }{ (N+j+3)! }. 
\end{equation}
The latter identity can be proved with mathematical induction. 
The evaluation of the summation on the r.h.s. of 
Eq.~(\ref{eq:summation_by_parts}) reduces now to sums of the form 
\begin{equation}\label{eq:F_nu_m_n}
    F^s_{m,n} = \sum_{j=0}^n \frac{ (-1)^j j^s }{ (j+m)! } ,
\end{equation}
where $m$ and $s$ are non-negative integers. For $s=0$  
    $F^0_{n,m} = (-1)^m  ( d_{n+m} - d_{m-1} )$ 
while for $s >0$ one easily obtains the recurrence relation
\begin{equation}
    F^s_{m,n} = F^{s-1}_{m-1,n} - m F^{s-1}_{m,n} .
\end{equation}
The latter relation allows us to express $p_N$ in the closed form as
\begin{equation}
    p_N = \frac{ (-1)^N N! }{ (2N+1)! } - d_N \frac{ (N!)^2 }{ (2N)! } .
\end{equation}
Therefore the $K \rightarrow \infty$ limit of $q_N^{(K)}$, denoted 
by $q_N$, is 
\begin{equation}\label{eq:q_N_final}
    q_N = -4 d_N^2 (3N+4) \frac{ [(N+1)!]^2 }{ (2N+3)! } + 2 d_N \frac{ (-1)^N N! }{ (2N+1)! } ,
\end{equation}
and consequently
\begin{equation}\label{ratio}
    \rho_N ( j_0^{\textrm{var}} ) = 4 - 2 N^{-1} + O(N^{-2}) , 
\end{equation}
in agreement with the numerical results of Ref.~\cite{Gniewek:15}. 
Thus, the series of approximations to the exchange energy
$j_0$ obtained from the multipole expansion of the primitive function 
and the variational formula converges at large $N$ 
like a geometric series with the term ratio equal to 4. 
In Section 4 we shall show that, surprisingly enough, this 
convergence changes 
to harmonic when $K$ is finite, i.e., when $\Phi$ is replaced with 
its approximation by a finite-order polarization function $\Phi_N^{(K)}$.

\subsubsection{Partial sums for the variational formula}

Recalling the definition of $q_N$ as the numerator of 
Eq. (\ref{eq:rho_K_N_j0surf}) times $2/e$ one can easily see 
that the constant $j_0$  calculated from the variational formula 
with the multipole expansion of the primitive function through 
the $1/R^N$ terms can be obtained by summing the $q_j$ increments 
up to $j=N$. Using 
Eq.~(\ref{eq:q_N_final}) one finds  
\begin{equation}\label{eq:j0_var_start}
    \frac{2}{e} j_0^{\textrm{var}} [ \Phi_N ] = \sum_{j=0}^N q_j 
    = Q_N +L_N, 
\end{equation}
where  the $j$ summation in Eq.~(\ref{eq:j0_var_start}) was separated 
into the $Q_N$ contribution containing terms quadratic in $d_j^2$ 
and the $L_N$ part 
linear in $d_j$. 
Both these sums can be calculated using the summation by parts formula of 
Eq.~(\ref{eq:summation_by_parts}).
The $Q_N$ sum is obtained from this formula by setting $u_j = -4 d_j^2$ and  
$w_j = (3j+4)[(j+1)! ]^2 /(2j+3)!$ 
The differences of  $u_j$'s are then given by  
$ u_{j+1} -  u_j = \alpha_j + \beta_j$,   
where
\begin{equation}
    \alpha_j = 8\, d_j \,\frac{ (-1)^j }{ (j+1)! } , \quad \quad  
    \beta_j = -4 \,\frac{ 1 }{ [ (j+1)! ]^2 } .
\end{equation}
The sums  $w_0+\cdots + w_j$ can be proved with the mathematical 
induction to be  equal to $W_j=1+\gamma_j$, where 
\begin{equation}
    \gamma_j = -2 \,\frac{ [ (j+2)! ]^2 }{ (2j+4)! } . 
\end{equation}
The summation formula of Eq.~(\ref{eq:summation_by_parts})
gives now 
\begin{equation}\label{eq:S1_final}
    Q_N = -4 -4\,d^2_N\,\gamma_N
  - \sum_{j=0}^{N-1} ( \alpha_j + \beta_j )\, \gamma_j .
\end{equation}
Using the definitions of $\alpha_j$ and $\gamma_j$  the $L_N$ term 
in Eq.~(\ref{eq:j0_var_start}) can be written in the form
\begin{equation}\label{eq:S2_modified}
    L_N = \sum_{j=0}^{N-1} u_j w_j  + \sum_{j=0}^{N-1} \alpha_j \gamma_j 
 + 2 d_N \frac{ (-1)^N N! }{ (2N+1)! } 
\end{equation}
where now we have set $u_j = d_j$ and 
\begin{equation}
    w_j = 2 \frac{ 4j+7 }{ 2j+3 } \frac{ (-1)^j j! }{ (2j+1)! } .
\end{equation}
Note that the sums containing  $\alpha_j\gamma_j$ cancel out when the sum 
$Q_N+L_N$ is taken. 
To execute the first sum on the r.h.s. of Eq.~(\ref{eq:S2_modified}) 
using the  summation by parts formula we need an expression 
for the sums $W_j=w_0+\cdots + w_j$.  
With the help of the mathematical induction we find that  
$W_j= 4 + \delta_j$, where  
\begin{equation}
    \delta_j = 4 (-1)^j \frac{ (j+1)! }{ (2j+3)! } .
\end{equation}
Application of the summation by parts formula to the first term in 
Eq~(\ref{eq:S2_modified})  gives 
\begin{equation}\label{eq:uj_wj_prime_final}
    \sum_{j=0}^{N-1} u_j w_j = 4 + u_{N-1} \delta_{N-1} - \sum_{j=0}^{N-2} ( u_{j+1} - u_j ) \delta_j .
\end{equation}
Now the only contributions to $j_0^{\textrm{var}}[\Phi_N]$ that 
still have to summed are the sum of $\beta_j \gamma_j$ 
in Eq.~(\ref{eq:S1_final}) and the sum on the r.h.s. 
of (\ref{eq:uj_wj_prime_final}). These sums can be combined to yield 
\begin{equation}
    \sum_{j=0}^{N-2} \big [ \beta_j \gamma_j + ( u_{j+1} - u_j ) \delta_j \big ] = 2 d_{2N-1} .
\end{equation}
Using this result we obtain the desired closed form expression  
\begin{equation}
\frac{2}{e} j_0^{\textrm{var}} [ \Phi_N ] = - 2 d_{2N} + \frac{ 2 }{ (2N+1)! } 
- 2\frac{(-1)^N N! }{ (2N+1)! } d_N + 8 \frac{ [ (N+2)! ]^2 }{ (2N+4)! } d_N^2 .
\end{equation}
It is readily seen that the $N\rightarrow\infty$ limit of this formula, 
resulting from the first term on the r.h.s., is equal to $-2/e$ in agreement 
with the exact value $j_0=-1$. 
After some manipulations one can also verify that the corresponding 
increment ratio of Eq. (\ref{eq:rho_K_N_j0surf}) 
agrees with the estimate of Eq. (\ref{ratio}).    

\section{Convergence rate in the case of truncated 
  polarization expansion}

\subsection{The surface-integral formula}

When $K$ is finite the increment ratio of 
Eq. (\ref{eq:rho_K_N_j0surf}) takes the form 
\begin{equation}\label{eq:rho_j0_surf_A_B}
    \rho_N^{(K)} ( j^{\textrm{surf}}_0 ) = A_N^{(K)} B_N^{(K)} ,
\end{equation}
where $ A_N^{(K)} =  d_N^{(K)} / d_{N+1}^{(K)} $
and
\begin{equation}\label{eq:B_K_N_j0surf}
    B_N^{(K)} = 2 \frac{ r_N^{(K)} - 2^{-N-1} d_N^{(K)} }{ r_N^{(K)} + 2^{-N-2} d_{N+1}^{(K)} } .
\end{equation}
Equation (\ref{eq:B_K_N_j0surf}) may be regarded  a finite $K$ precursor
of Eq. (\ref{eq:rho_j0_surf_K_infty}).
In view of Eq. (\ref{eq:d_K_N_asymptotics}) 
the asymptotics of $A_N^{(K)}$ is given by 
\begin{equation}\label{eq:A_K_N_asymptotics}
    A_N^{(K)} = 1 + N^{-1} - ( K-1 ) N^{-1} ( \ln N )^{-1} + O \big ( N^{-1} ( \ln N )^{-2} \big ) .
\end{equation}
After  expanding Eq.~(\ref{eq:B_K_N_j0surf}) one obtains the following 
asymptotic expression for $B_N^{(K)}$ 
\begin{equation}\label{eq:B_K_N_asymptotics}
    B_N^{(K)} = 2 - c_N^{(K)} 2^{-N} + O(4^{-N}) ,
\end{equation}
where 
\begin{equation}\label{eq:c_K_N_definition}
    c_N^{(K)} =  \frac{ 2 d_N^{(K)} + d_{N+1}^{(K)} }{ 2 r_N^{(K)} } .
\end{equation}
Equation~(\ref{eq:rNK_def}) shows that $r_N^{(K)}$ is a sum of rapidly decaying 
terms, and therefore converges quickly to its limit. Specifically it can be showed that 
for $N > 2K+1$ 
\begin{equation}
    r_{2K+1} \leq r_N^{(K)} \leq r_N
\end{equation}
and
\begin{equation}
    r_N^{(K)} = 2 e_{2K+1}(-\tfrac{1}{2}) + O(2^{-2K}) .
\end{equation}

For this reason asymptotics of $c_N^{(K)}$ is determined by the large $N$ behavior of 
$d_N^{(K)}$. 
Eq.~(\ref{eq:d_K_N_asymptotics}) implies then that  
at large $N$ and fixed $K$ we have
\begin{equation}
    c_N^{(K)} = \frac{ 3 ( \ln N )^{K-1} }{ 4 e_{2K+1}(-\tfrac{1}{2}) (K-1)! N } \big [ 1 + O(( \ln N )^{-1}) \big ] . 
\end{equation}
The $N \rightarrow \infty$ limit of $c_N^{(K)}$ (with fixed $K$) is therefore 
\begin{equation}
    \lim_{N \rightarrow \infty} c_N^{(K)} = 0 .
\end{equation}
On the other hand the $K \rightarrow \infty$ limit of $c_N^{(K)}$ for a fixed $N$ is
\begin{equation}
    c_N = \lim_{K \rightarrow \infty} c_N^{(K)} = \frac{ 3 d_N }{ 2 r_N } \big [ 1 + O((N!)^{-1}) \big ] .
\end{equation}
Ultimately the limit of $c_N$ is 
\begin{equation}\label{eq:cN_limit}
    \lim_{ N \rightarrow \infty } c_N = \frac{ 3 }{ 4\sqrt{e} } ,
\end{equation}
in agreement with Eq.~(\ref{eq:rho_N_j0surf}). 
Note that the combined limit $K \rightarrow \infty, N \rightarrow \infty$ of 
$c_N^{(K)}$ does not exist, as the result depends on the order of limits.

The asymptotics of the increment ratio of 
$j_0^{\textrm{surf}} [ \Phi_N^{(K)}]$ is given by
\begin{equation}\label{rhoKNsurf}
    \rho_N^{(K)} ( j_0^{\textrm{surf}} ) = 2 + 2 N^{-1} - 2(K-1) N^{-1} ( \ln N )^{-1} + O \big ( N^{-1} ( \ln N )^{-2} \big ) .
\end{equation}
Eq. (\ref{rhoKNsurf}) agrees with the results presented for $K=1$ 
in Ref.~\cite{Gniewek:15}. This equation shows that for finite 
$K$ the convergence radius corresponding 
to $\rho_N^{(K)} ( j_0^{\textrm{surf}} )$ and equal 2 
is the same as for $K =\infty$. 
It may seem disturbing that the $K\rightarrow \infty$ limit 
of Eq.~(\ref{rhoKNsurf}) does not coincide with Eq.~(\ref{eq:rho_N_j0surf}).  
This apparent inconsistency results from the facts that the 
$K \rightarrow \infty$ and $N \rightarrow \infty$ limits of $\rho_N^{(K)} ( j_0^{\textrm{surf}} )$  do not commute. 

\subsection{The variational formula}

In this subsection we shall show that for finite $K$ the convergence of of 
 $\rho_N^{(K)} ( j_0^{\textrm{var}} )$ corresponds to the convergence radius 
of 1, i.e. becomes harmonic, in contrast to the  fast geometric convergence 
found for $K=\infty$. 

\begin{figure}
\begin{center}
\begin{minipage}{100mm}
\resizebox*{10cm}{!}{\includegraphics{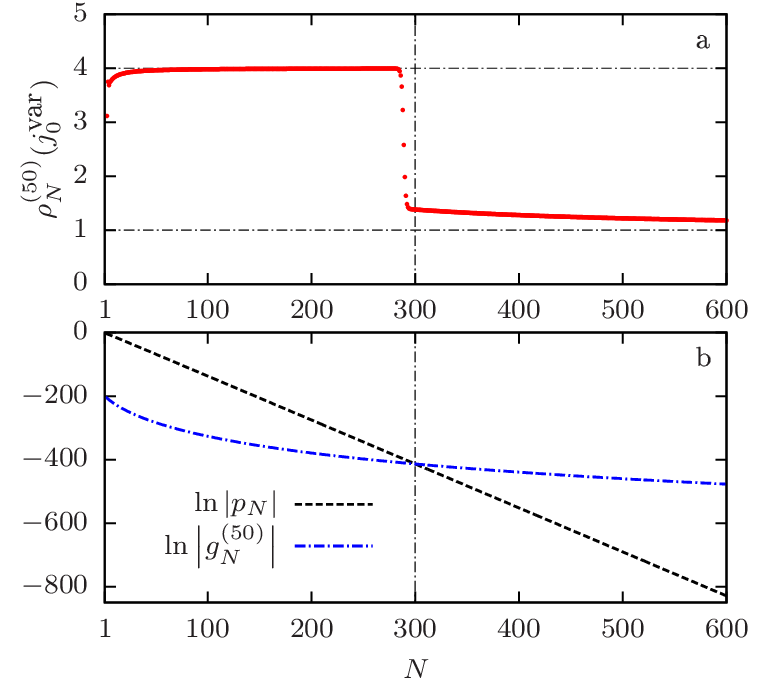}}%
\caption{Switching of convergence rate in $\rho_N^{(50)} (j_0^{(\textrm{var})} )$.  }%
\label{fig:rho_var}
\end{minipage}
\end{center}
\end{figure}

The inverse d'Alembert ratio $\rho_N^{(K)} ( j_0^{\textrm{var}} )$ can be factored as 
\begin{equation}\label{eq:rho_j0var_factored}
    \rho_N^{(K)} ( j_0^{\textrm{var}} ) = A_N^{(K)} D_N^{(K)} , 
\end{equation}
similarly to Eq.~(\ref{eq:rho_j0_surf_A_B}) for the surface-integral formula. 
$D_N^{(K)}$ is defined as
\begin{equation}
    D_N^{(K)} = \frac{ 2 p_N^{(K)} + d_N^{(K)} G_{N,N} }{ 2 p_{N+1}^{(K)} + d_{N+1}^{(K)} G_{N+1,N+1} } .
\end{equation}
As $d_N^{(K)} = d_N$ for $N \leq 2K+1$ we have
\begin{equation}\label{eq:p_K_N_minus_p_N}
    p_N^{(K)} - p_N = \sum_{m=2K+2}^{N-1} \big ( d_m^{(K)} - d_m \big ) G_{m,N} .
\end{equation}
Since 
\begin{equation}
    \frac{ G_{m,N} }{ G_{m+1,N} } = \frac{ m N }{ (m+1)^2 } \big [ 1 + O(N^{-1}) \big ] 
\end{equation}
the $m=2K+2$ term asymptotically dominates the sum in Eq.~(\ref{eq:p_K_N_minus_p_N}),
\begin{equation}
    p_N^{(K)} - p_N = g_N^{(K)} \big [ 1 + O(N^{-1}) \big ] ,
\end{equation}
where we have defined 
\begin{equation}
    g_N^{(K)} = \big ( d_{2K+2}^{(K)} - d_{2K+2} \big ) G_{2K+2,N} .
\end{equation}
Comparison with Eq.~(\ref{eq:d_N_short_sum}) shows that 
\begin{equation}
    g_N^{(K)} = - t_{2K+2}^{(K+1)} G_{2K+2,N} .
\end{equation}
From Eq.~(\ref{eq:t_k_n_recurrence}) we see  that 
\begin{equation}
    t_{2k}^{(k)} = \frac{1}{2k} t_{2k-2}^{(k-1)} = \frac{ 1 }{ (2k)!! } ,
\end{equation}
therefore
\begin{equation}
    g_N^{(K)} = - \frac{ G_{2K+2,N} }{ (2K+2)!! } .
\end{equation}
Stirling's approximation shows that $G_{N,N} = O(\sqrt{N} 2^{-2N})$, 
thus
\begin{equation}
    D_N^{(K)} = \frac{ p_N - g_N^{(K)} }{ p_{N+1} - g_{N+1}^{(K)} } \big [ 1 + O(N^{-1}) \big ] .
\end{equation}
The switching of the convergence rate of
 $j_0^{\textrm{var}} [ \Phi_N^{(K)} ]$ is caused by the change of relative importance of 
$p_N$ and $g_N^{(K)}$, as the ratios of consecutive elements of these sequences are  
\begin{equation}
    \frac{ p_N }{ p_{N+1} } = 4 - 2 N^{-1} + O(N^{-2}) ,
\end{equation}
and 
\begin{equation}\label{eq:gNK_ratio}
    \frac{ g_N^{(K)} }{ g_{N+1}^{(K)} } = 1 + (2K+3) N^{-1} + O( N^{-2} ) .
\end{equation}
This switching phenomenon is illustrated in Fig.~\ref{fig:rho_var}.

The change of convergence pattern, observed numerically 
in Ref. \cite{Gniewek:15}, takes place for $N = N_c$, where $N_c$ is given as a solution of 
$| p_{N_c} | \approx | g_{N_c}^{(K)} |$, which means that asymptotically 
\begin{equation}\label{eq:new_Nc_eq}
    \frac{ \Gamma( N_c ) \Gamma( N_c + 2K + 5 ) }{ \Gamma( 2N_c + 1 ) } = 2 e \frac{ (2K+2)! }{ (2K)!! } .
\end{equation}
Equation~(\ref{eq:new_Nc_eq}) is slightly different from Eq.~(59) of our previous work, Ref.~\cite{Gniewek:15}, 
which contains a numerical error. However, the solutions $N_c$ 
of both equations  are asymptotically equal 
for large $K$. These solutions 
differ by less than 0.5\% already for $K=40$ 
and agree very well with the numerically observed values \cite{Gniewek:15}. 
Stirling's approximation applied to both sides of Eq.~(\ref{eq:new_Nc_eq}) shows that asymptotically 
\begin{equation}
    N_c \sim K \ln K .
\end{equation}
Equations~(\ref{eq:rho_j0var_factored}) and (\ref{eq:gNK_ratio})
show that for $N > N_c$ the increment ratio for 
$j_0^{\textrm{var}} [ \Phi_N^{(K)} ]$ behaves as
\begin{equation}\label{rhoKNvar}
    \rho_N^{(K)}( j_0^{\textrm{var}} ) = 1 + (2K+4)N^{-1} + (K-1) N^{-1} (\ln N)^{-1} + O( N^{-1} (\ln N)^{-2} ) ,
\end{equation}
confirming the numerical results of Ref.~\cite{Gniewek:15}. The fact that 
the $K\rightarrow \infty$ limit
of Eq.~(\ref{rhoKNvar}) does not coincide with Eq.~(\ref{ratio})
results from the fact that the 
$K \rightarrow \infty$ and $N \rightarrow \infty$ limits of $\rho_N^{(K)} ( j_0^{\textrm{var}} )$  do not commute.

\section{Summary and conclusions}

Using the molecular hydrogen ion as a model system we have analytically 
investigated the applicability of the conventional multipole 
expansion of the wave function to obtain the large-$R$ 
asymptotics of the exchange contribution to the interaction energy. 
We considered the well-known surface integral formula as well as the 
volume-integral formula derived from the variational principle.  
We have obtained closed-form expressions for partial sums of
the corresponding sequences of approximations to the constant 
$j_0$ characterizing the exponential fall off of the exchange energy.
Furthermore, we obtained analytical expressions for the large-$N$ behavior,
$N$ being the maximal oder of the multiple expansion included, 
of the increment rations of the considered approximations to $j_0$. 
These partial sums converge to the correct limit $j_0 =-1$ for both 
investigated formulas (for the conventional  SAPT formula this was 
proven in Ref. \cite{Gniewek:15}).
The variational formula is characterized by a remarkably fast convergence, 
matching that of a geometric series with the quotient $ = \tfrac{1}{4}$ 
(or, equivalently, having the convergence radius $\rho = 4$). 
This convergence is much faster than that resulting from the application 
of the conventional SAPT formula  \cite{Gniewek:15}.
In the case of the surface-integral formula we observed slightly slower convergence, 
namely that of a geometric series with the quotient $q = \tfrac{1}{2}$ 
(or $\rho = 2$). 
We have also demonstrated a peculiar switching from geometric to 
harmonic convergence in the case of calculations based on the variational 
formula and the multipole expansion of a truncated polarization 
function. In this case for $N < N_c$, where $N_c$ is a certain critical
value of $N$,  the increment ratios behave 
like $4 - 2 N^{-1}$, whereas for $N > N_c$ its leading term changes 
to unity and the sub-leading term depends on the order of the 
polarization function, $K$. 
The critical value $N_c$ is roughly five or six times greater 
than $K$ for $K$ less than 100 
and grows as $ K \ln K $ for large $K$.
In the case of the surface-integral formula 
we do not observe a similar change of the convergence rate when
a finite-order polarization expansion is used to approximate $\Phi$.   

Our analytic results show that the variational volume integral formula 
offers a considerable advantage, compared to the standard SAPT formula, 
in computing the exchange contribution to the interaction energy 
using the conventional multpole and/or basis set expansion 
of the wave function. This observation suggest a possibility of a 
significant improvement over the conventional SAPT approach 
to the problem of molecular interactions, both in terms of 
the convergence with respect to the order of the interaction potential 
and the basis set size. 

\section*{Acknowledgment}

The authors thank  Robert Moszy{\'n}ski for commenting on the
manuscript. 
This work was supported by the 
National Science Centre, Poland, 
project number 2014/13/N/ST4/03833.

\bibliography{ref}

\begin{thebibliography}{41}
\expandafter\ifx\csname natexlab\endcsname\relax\def\natexlab#1{#1}\fi
\expandafter\ifx\csname bibnamefont\endcsname\relax
  \def\bibnamefont#1{#1}\fi
\expandafter\ifx\csname bibfnamefont\endcsname\relax
  \def\bibfnamefont#1{#1}\fi
\expandafter\ifx\csname citenamefont\endcsname\relax
  \def\citenamefont#1{#1}\fi
\expandafter\ifx\csname url\endcsname\relax
  \def\url#1{\texttt{#1}}\fi
\expandafter\ifx\csname urlprefix\endcsname\relax\def\urlprefix{URL }\fi
\providecommand{\bibinfo}[2]{#2}
\providecommand{\eprint}[2][]{\url{#2}}

\bibitem[{\citenamefont{Kaplan}(2006)}]{Kaplan:06}
\bibinfo{author}{\bibfnamefont{I.~G.} \bibnamefont{Kaplan}}, in
  \emph{\bibinfo{booktitle}{Intermolecular Interactions: Physical Picture,
  Computational Methods and Model Potentials}} (\bibinfo{publisher}{John Wiley
  \& Sons}, \bibinfo{address}{Chichester}, \bibinfo{year}{2006}).

\bibitem[{\citenamefont{Stone}(2013)}]{Stone:13}
\bibinfo{author}{\bibfnamefont{A.~J.} \bibnamefont{Stone}}, in
  \emph{\bibinfo{booktitle}{The Theory of Intermolecular Forces}}
  (\bibinfo{publisher}{University Press}, \bibinfo{address}{Oxford},
  \bibinfo{year}{2013}).

\bibitem[{\citenamefont{Kim et~al.}(2015)\citenamefont{Kim, Weichman,
  Sjolander, Neumark, K{\l}os, Alexander, and Manolopoulos}}]{Kim:15}
\bibinfo{author}{\bibfnamefont{J.~B.} \bibnamefont{Kim}},
  \bibinfo{author}{\bibfnamefont{M.~L.} \bibnamefont{Weichman}},
  \bibinfo{author}{\bibfnamefont{T.~F.} \bibnamefont{Sjolander}},
  \bibinfo{author}{\bibfnamefont{D.~M.} \bibnamefont{Neumark}},
  \bibinfo{author}{\bibfnamefont{J.}~\bibnamefont{K{\l}os}},
  \bibinfo{author}{\bibfnamefont{M.~H.} \bibnamefont{Alexander}},
  \bibnamefont{and} \bibinfo{author}{\bibfnamefont{D.~E.}
  \bibnamefont{Manolopoulos}}, \bibinfo{journal}{Science}
  \textbf{\bibinfo{volume}{349}}, \bibinfo{pages}{510} (\bibinfo{year}{2015}).

\bibitem[{\citenamefont{Jeziorski et~al.}(1994)\citenamefont{Jeziorski,
  Moszy\'nski, and Szalewicz}}]{Jeziorski:94}
\bibinfo{author}{\bibfnamefont{B.}~\bibnamefont{Jeziorski}},
  \bibinfo{author}{\bibfnamefont{R.}~\bibnamefont{Moszy\'nski}},
  \bibnamefont{and}
  \bibinfo{author}{\bibfnamefont{K.}~\bibnamefont{Szalewicz}},
  \bibinfo{journal}{Chem. Rev.} \textbf{\bibinfo{volume}{94}},
  \bibinfo{pages}{1887} (\bibinfo{year}{1994}).

\bibitem[{\citenamefont{Szalewicz et~al.}(2005)\citenamefont{Szalewicz,
  Patkowski, and Jeziorski}}]{Szalewicz:05}
\bibinfo{author}{\bibfnamefont{K.}~\bibnamefont{Szalewicz}},
  \bibinfo{author}{\bibfnamefont{K.}~\bibnamefont{Patkowski}},
  \bibnamefont{and}
  \bibinfo{author}{\bibfnamefont{B.}~\bibnamefont{Jeziorski}}, in
  \emph{\bibinfo{booktitle}{Intermolecular Forces and Clusters (Structure and
  Bonding, volume 116)}}, edited by \bibinfo{editor}{\bibfnamefont{D.~J.}
  \bibnamefont{Wales}} (\bibinfo{publisher}{Springer-Verlag},
  \bibinfo{address}{Heidelberg}, \bibinfo{year}{2005}), pp.
  \bibinfo{pages}{43--117}.

\bibitem[{\citenamefont{Young}(1975)}]{Young:75}
\bibinfo{author}{\bibfnamefont{R.~H.} \bibnamefont{Young}},
  \bibinfo{journal}{Int. J. Quant. Chem.} \textbf{\bibinfo{volume}{9}},
  \bibinfo{pages}{47} (\bibinfo{year}{1975}).

\bibitem[{\citenamefont{{J. \v{C}\'\i\v{z}ek, R. J. Damburg, S. Graffi, V.
  Grecchi, E. M. Harrell II, J. G. Harris, S. Nakai, J. Paldus, R. Kh. Propin,
  and H. J. Silverstone}}(1986)}]{Cizek:86}
\bibinfo{author}{\bibnamefont{{J. \v{C}\'\i\v{z}ek, R. J. Damburg, S. Graffi,
  V. Grecchi, E. M. Harrell II, J. G. Harris, S. Nakai, J. Paldus, R. Kh.
  Propin, and H. J. Silverstone}}}, \bibinfo{journal}{Phys. Rev. A}
  \textbf{\bibinfo{volume}{33}}, \bibinfo{pages}{12} (\bibinfo{year}{1986}).

\bibitem[{\citenamefont{Ahlrichs}(1976)}]{Ahlrichs:76}
\bibinfo{author}{\bibfnamefont{R.}~\bibnamefont{Ahlrichs}},
  \bibinfo{journal}{Theor. Chim. Acta} \textbf{\bibinfo{volume}{41}},
  \bibinfo{pages}{7} (\bibinfo{year}{1976}).

\bibitem[{\citenamefont{{J. D. Morgan III and B. Simon}}(1980)}]{Morgan:80}
\bibinfo{author}{\bibnamefont{{J. D. Morgan III and B. Simon}}},
  \bibinfo{journal}{Int. J. Quant. Chem.} \textbf{\bibinfo{volume}{17}},
  \bibinfo{pages}{1143} (\bibinfo{year}{1980}).

\bibitem[{\citenamefont{Cha{\l}asi\'nski
  et~al.}(1977)\citenamefont{Cha{\l}asi\'nski, Jeziorski, and
  Szalewicz}}]{Chalasinski:77}
\bibinfo{author}{\bibfnamefont{G.}~\bibnamefont{Cha{\l}asi\'nski}},
  \bibinfo{author}{\bibfnamefont{B.}~\bibnamefont{Jeziorski}},
  \bibnamefont{and}
  \bibinfo{author}{\bibfnamefont{K.}~\bibnamefont{Szalewicz}},
  \bibinfo{journal}{Int. J. Quant. Chem.} \textbf{\bibinfo{volume}{11}},
  \bibinfo{pages}{247} (\bibinfo{year}{1977}).

\bibitem[{\citenamefont{\'Cwiok et~al.}(1992)\citenamefont{\'Cwiok, Jeziorski,
  Ko\l{}os, Moszy\'nski, Rychlewski, and Szalewicz}}]{Cwiok:92:Pol}
\bibinfo{author}{\bibfnamefont{T.}~\bibnamefont{\'Cwiok}},
  \bibinfo{author}{\bibfnamefont{B.}~\bibnamefont{Jeziorski}},
  \bibinfo{author}{\bibfnamefont{W.}~\bibnamefont{Ko\l{}os}},
  \bibinfo{author}{\bibfnamefont{R.}~\bibnamefont{Moszy\'nski}},
  \bibinfo{author}{\bibfnamefont{J.}~\bibnamefont{Rychlewski}},
  \bibnamefont{and}
  \bibinfo{author}{\bibfnamefont{K.}~\bibnamefont{Szalewicz}},
  \bibinfo{journal}{Chem. Phys. Lett.} \textbf{\bibinfo{volume}{195}},
  \bibinfo{pages}{67} (\bibinfo{year}{1992}).

\bibitem[{\citenamefont{Jeziorski and Ko\l{}os}(1977)}]{Jeziorski:77}
\bibinfo{author}{\bibfnamefont{B.}~\bibnamefont{Jeziorski}} \bibnamefont{and}
  \bibinfo{author}{\bibfnamefont{W.}~\bibnamefont{Ko\l{}os}},
  \bibinfo{journal}{Int. J. Quant. Chem.} \textbf{\bibinfo{volume}{S12}},
  \bibinfo{pages}{91} (\bibinfo{year}{1977}).

\bibitem[{\citenamefont{Kutzelnigg}(1978)}]{Kutzelnigg:78}
\bibinfo{author}{\bibfnamefont{W.}~\bibnamefont{Kutzelnigg}},
  \bibinfo{journal}{Int. J. Quant. Chem.} \textbf{\bibinfo{volume}{14}},
  \bibinfo{pages}{101} (\bibinfo{year}{1978}).

\bibitem[{\citenamefont{Jeziorski et~al.}(1978)\citenamefont{Jeziorski,
  Szalewicz, and Cha\l{}asi\'nski}}]{Jeziorski:78}
\bibinfo{author}{\bibfnamefont{B.}~\bibnamefont{Jeziorski}},
  \bibinfo{author}{\bibfnamefont{K.}~\bibnamefont{Szalewicz}},
  \bibnamefont{and}
  \bibinfo{author}{\bibfnamefont{G.}~\bibnamefont{Cha\l{}asi\'nski}},
  \bibinfo{journal}{Int. J. Quant. Chem.} \textbf{\bibinfo{volume}{14}},
  \bibinfo{pages}{271} (\bibinfo{year}{1978}).

\bibitem[{\citenamefont{Szalewicz}(2012)}]{Szalewicz:12}
\bibinfo{author}{\bibfnamefont{K.}~\bibnamefont{Szalewicz}},
  \bibinfo{journal}{WIREs: Comput. Mol. Sci.} \textbf{\bibinfo{volume}{2}},
  \bibinfo{pages}{254} (\bibinfo{year}{2012}).

\bibitem[{\citenamefont{Hohenstein and Sherrill}(2012)}]{Sherrill:12}
\bibinfo{author}{\bibfnamefont{E.~G.} \bibnamefont{Hohenstein}}
  \bibnamefont{and} \bibinfo{author}{\bibfnamefont{C.~D.}
  \bibnamefont{Sherrill}}, \bibinfo{journal}{WIREs: Comput. Mol. Sci.}
  \textbf{\bibinfo{volume}{2}}, \bibinfo{pages}{304} (\bibinfo{year}{2012}).

\bibitem[{\citenamefont{Jansen}(2014)}]{Jansen:14}
\bibinfo{author}{\bibfnamefont{G.}~\bibnamefont{Jansen}},
  \bibinfo{journal}{WIREs: Comput. Mol. Sci.} \textbf{\bibinfo{volume}{4}},
  \bibinfo{pages}{127} (\bibinfo{year}{2014}).

\bibitem[{\citenamefont{Hesselmann and Korona}(2014)}]{Hesselmann:14}
\bibinfo{author}{\bibfnamefont{A.}~\bibnamefont{Hesselmann}} \bibnamefont{and}
  \bibinfo{author}{\bibfnamefont{T.}~\bibnamefont{Korona}},
  \bibinfo{journal}{J. Chem. Phys.} \textbf{\bibinfo{volume}{141}},
  \bibinfo{pages}{094107} (\bibinfo{year}{2014}).

\bibitem[{\citenamefont{Kutzelnigg}(1980)}]{Kutzelnigg:80}
\bibinfo{author}{\bibfnamefont{W.}~\bibnamefont{Kutzelnigg}},
  \bibinfo{journal}{J. Chem. Phys.} \textbf{\bibinfo{volume}{73}},
  \bibinfo{pages}{343} (\bibinfo{year}{1980}).

\bibitem[{\citenamefont{Adams}(1996)}]{Adams:96}
\bibinfo{author}{\bibfnamefont{W.~H.} \bibnamefont{Adams}},
  \bibinfo{journal}{Int. J. Quant. Chem.} \textbf{\bibinfo{volume}{60}},
  \bibinfo{pages}{273} (\bibinfo{year}{1996}).

\bibitem[{\citenamefont{Patkowski et~al.}(2001)\citenamefont{Patkowski, Korona,
  and Jeziorski}}]{Patkowski:01}
\bibinfo{author}{\bibfnamefont{K.}~\bibnamefont{Patkowski}},
  \bibinfo{author}{\bibfnamefont{T.}~\bibnamefont{Korona}}, \bibnamefont{and}
  \bibinfo{author}{\bibfnamefont{B.}~\bibnamefont{Jeziorski}},
  \bibinfo{journal}{J. Chem. Phys.} \textbf{\bibinfo{volume}{115}},
  \bibinfo{pages}{1137} (\bibinfo{year}{2001}).

\bibitem[{\citenamefont{Przybytek et~al.}(2004)\citenamefont{Przybytek,
  Patkowski, and Jeziorski}}]{Przybytek:04}
\bibinfo{author}{\bibfnamefont{M.}~\bibnamefont{Przybytek}},
  \bibinfo{author}{\bibfnamefont{K.}~\bibnamefont{Patkowski}},
  \bibnamefont{and}
  \bibinfo{author}{\bibfnamefont{B.}~\bibnamefont{Jeziorski}},
  \textbf{\bibinfo{volume}{69}}, \bibinfo{pages}{141} (\bibinfo{year}{2004}).

\bibitem[{\citenamefont{Kutzelnigg}(1992)}]{Kutzelnigg:92}
\bibinfo{author}{\bibfnamefont{W.}~\bibnamefont{Kutzelnigg}},
  \bibinfo{journal}{Chem. Phys. Lett.} \textbf{\bibinfo{volume}{195}},
  \bibinfo{pages}{77 } (\bibinfo{year}{1992}).

\bibitem[{\citenamefont{Jeziorski et~al.}(1980)\citenamefont{Jeziorski,
  Schwalm, and Szalewicz}}]{JeziorskiSchwalmSzalewicz1980}
\bibinfo{author}{\bibfnamefont{B.}~\bibnamefont{Jeziorski}},
  \bibinfo{author}{\bibfnamefont{W.~A.} \bibnamefont{Schwalm}},
  \bibnamefont{and}
  \bibinfo{author}{\bibfnamefont{K.}~\bibnamefont{Szalewicz}},
  \bibinfo{journal}{J. Chem. Phys.} \textbf{\bibinfo{volume}{73}},
  \bibinfo{pages}{6215} (\bibinfo{year}{1980}).

\bibitem[{\citenamefont{Gniewek and Jeziorski}(2014)}]{Gniewek:14}
\bibinfo{author}{\bibfnamefont{P.}~\bibnamefont{Gniewek}} \bibnamefont{and}
  \bibinfo{author}{\bibfnamefont{B.}~\bibnamefont{Jeziorski}},
  \bibinfo{journal}{Phys. Rev. A} \textbf{\bibinfo{volume}{90}},
  \bibinfo{pages}{022506} (\bibinfo{year}{2014}).

\bibitem[{\citenamefont{Jeziorski and Ko{\l}os}(1982)}]{Jeziorski:82}
\bibinfo{author}{\bibfnamefont{B.}~\bibnamefont{Jeziorski}} \bibnamefont{and}
  \bibinfo{author}{\bibfnamefont{W.}~\bibnamefont{Ko{\l}os}}, in
  \emph{\bibinfo{booktitle}{Molecular interactions}}, edited by
  \bibinfo{editor}{\bibfnamefont{H.}~\bibnamefont{Ratajczak}} \bibnamefont{and}
  \bibinfo{editor}{\bibfnamefont{W.~J.} \bibnamefont{Orville-Thomas}}
  (\bibinfo{publisher}{Wiley}, \bibinfo{address}{New York},
  \bibinfo{year}{1982}), vol.~\bibinfo{volume}{3}, pp. \bibinfo{pages}{1--46}.

\bibitem[{\citenamefont{Chipman and Hirschfelder}(1973)}]{Chipman:73}
\bibinfo{author}{\bibfnamefont{D.~M.} \bibnamefont{Chipman}} \bibnamefont{and}
  \bibinfo{author}{\bibfnamefont{J.~O.} \bibnamefont{Hirschfelder}},
  \bibinfo{journal}{J. Chem. Phys.} \textbf{\bibinfo{volume}{59}},
  \bibinfo{pages}{2838} (\bibinfo{year}{1973}).

\bibitem[{\citenamefont{Whitton and Byers-Brown}(1976)}]{Whitton:76}
\bibinfo{author}{\bibfnamefont{W.~N.} \bibnamefont{Whitton}} \bibnamefont{and}
  \bibinfo{author}{\bibfnamefont{W.}~\bibnamefont{Byers-Brown}},
  \bibinfo{journal}{Int. J. Quant. Chem.} \textbf{\bibinfo{volume}{10}},
  \bibinfo{pages}{71} (\bibinfo{year}{1976}).

\bibitem[{\citenamefont{{R. J. Damburg, R. Kh. Propin, S. Graffi, V. Grecchi,
  E. M. Harrell II, J. \v{C}\'\i\v{z}ek, J. Paldus, and H. J.
  Silverstone}}(1984)}]{Damburg:84}
\bibinfo{author}{\bibnamefont{{R. J. Damburg, R. Kh. Propin, S. Graffi, V.
  Grecchi, E. M. Harrell II, J. \v{C}\'\i\v{z}ek, J. Paldus, and H. J.
  Silverstone}}}, \bibinfo{journal}{Phys. Rev. Lett.}
  \textbf{\bibinfo{volume}{52}}, \bibinfo{pages}{1112} (\bibinfo{year}{1984}).

\bibitem[{\citenamefont{Harrell}(1980)}]{Harrell:80}
\bibinfo{author}{\bibfnamefont{E.~M.} \bibnamefont{Harrell}},
  \bibinfo{journal}{Commun. Math. Phys.} \textbf{\bibinfo{volume}{75}},
  \bibinfo{pages}{239} (\bibinfo{year}{1980}).

\bibitem[{\citenamefont{Damburg and Propin}(1982)}]{Damburg:82}
\bibinfo{author}{\bibfnamefont{R.}~\bibnamefont{Damburg}} \bibnamefont{and}
  \bibinfo{author}{\bibfnamefont{R.}~\bibnamefont{Propin}},
  \bibinfo{journal}{Int. J. Quant. Chem.} \textbf{\bibinfo{volume}{21}},
  \bibinfo{pages}{191} (\bibinfo{year}{1982}).

\bibitem[{\citenamefont{Firsov}(1951)}]{Firsov:51}
\bibinfo{author}{\bibfnamefont{O.~B.} \bibnamefont{Firsov}},
  \bibinfo{journal}{Zh. Eksp. Teor. Fiz.} \textbf{\bibinfo{volume}{21}},
  \bibinfo{pages}{1001} (\bibinfo{year}{1951}).

\bibitem[{\citenamefont{Holstein}(1952)}]{Holstein:52}
\bibinfo{author}{\bibfnamefont{T.}~\bibnamefont{Holstein}},
  \bibinfo{journal}{J. Phys. Chem.} \textbf{\bibinfo{volume}{56}},
  \bibinfo{pages}{832} (\bibinfo{year}{1952}).

\bibitem[{\citenamefont{Herring}(1962)}]{Herring:62}
\bibinfo{author}{\bibfnamefont{C.}~\bibnamefont{Herring}},
  \bibinfo{journal}{Rev. Mod. Phys.} \textbf{\bibinfo{volume}{34}},
  \bibinfo{pages}{631} (\bibinfo{year}{1962}).

\bibitem[{\citenamefont{Tang et~al.}(1998)\citenamefont{Tang, Toennies, and
  Yiu}}]{Tang:98}
\bibinfo{author}{\bibfnamefont{K.~T.} \bibnamefont{Tang}},
  \bibinfo{author}{\bibfnamefont{J.~P.} \bibnamefont{Toennies}},
  \bibnamefont{and} \bibinfo{author}{\bibfnamefont{C.~L.} \bibnamefont{Yiu}},
  \bibinfo{journal}{Int. Rev. Phys. Chem.} \textbf{\bibinfo{volume}{17}},
  \bibinfo{pages}{363} (\bibinfo{year}{1998}).

\bibitem[{\citenamefont{Gniewek and Jeziorski}(2015)}]{Gniewek:15}
\bibinfo{author}{\bibfnamefont{P.}~\bibnamefont{Gniewek}} \bibnamefont{and}
  \bibinfo{author}{\bibfnamefont{B.}~\bibnamefont{Jeziorski}},
  \bibinfo{journal}{J. Chem. Phys.} \textbf{\bibinfo{volume}{143}},
  \bibinfo{pages}{154106} (\bibinfo{year}{2015}).

\bibitem[{\citenamefont{Dalgarno and Lewis}(1955)}]{Dalgarno:55}
\bibinfo{author}{\bibfnamefont{A.}~\bibnamefont{Dalgarno}} \bibnamefont{and}
  \bibinfo{author}{\bibfnamefont{J.~T.} \bibnamefont{Lewis}},
  \bibinfo{journal}{Proc. Roy. Soc. A} \textbf{\bibinfo{volume}{233}},
  \bibinfo{pages}{70} (\bibinfo{year}{1955}).

\bibitem[{\citenamefont{Tang et~al.}(1991)\citenamefont{Tang, Toennies, and
  Yiu}}]{Tang:91}
\bibinfo{author}{\bibfnamefont{K.~T.} \bibnamefont{Tang}},
  \bibinfo{author}{\bibfnamefont{J.~P.} \bibnamefont{Toennies}},
  \bibnamefont{and} \bibinfo{author}{\bibfnamefont{C.~L.} \bibnamefont{Yiu}},
  \bibinfo{journal}{J. Chem. Phys.} \textbf{\bibinfo{volume}{94}},
  \bibinfo{pages}{7266} (\bibinfo{year}{1991}).

\bibitem[{\citenamefont{Riordan}(2014)}]{Riordan:14}
\bibinfo{author}{\bibfnamefont{J.}~\bibnamefont{Riordan}},
  \emph{\bibinfo{title}{An Introduction to Combinatorial Analysis}}
  (\bibinfo{publisher}{Princeton University Press}, \bibinfo{year}{2014}).

\bibitem[{\citenamefont{Comtet}(1974)}]{Comtet:74}
\bibinfo{author}{\bibfnamefont{L.}~\bibnamefont{Comtet}},
  \emph{\bibinfo{title}{Advanced Combinatorics: The art of finite and infinite
  expansions}} (\bibinfo{publisher}{Springer Science \& Business Media},
  \bibinfo{year}{1974}).

\bibitem[{OEI()}]{OEIS}
\bibinfo{note}{{OEIS Foundation Inc. (2011), The On-Line Encyclopedia of
  Integer Sequences, http://oeis.org}}.

\end{thebibliography}


\appendices

\section{\label{Appendix_tkn_asymptotics} Asymptotics of wave function coefficients}

In this Appendix we shall show that the wave function coefficients 
$t^{(k)}_n$ behave at large $n$ as 
\begin{equation}\label{eq:tkn_asymptotics}
    t_n^{(k)} = \frac{ (\ln n)^{k-1} }{ n (k-1)! }
 \bigg[ 1 + O \bigg( \frac{1}{\ln n} \bigg) \bigg] .
\end{equation}
For $k=1$ and $k=2$ the non-zero values of $t_n^{(k)}$ are
\begin{equation}\begin{aligned}
    t^{(1)}_n & = \frac{ 1 }{ n } ,  \quad  n \geq 2 ,  \\
    t^{(2)}_n & = \frac{ H_{n-2} }{ n } - \frac{ 1 }{ n } , \quad  n \geq 4 ,
\end{aligned}\end{equation}
where $H_n$ are the harmonic numbers 
\begin{equation}
    H_n = \sum_{k=1}^n \frac{ 1 }{ k } 
\end{equation}
growing asymptotically as $H_n = \ln n + \gamma +  O(n^{-1})$, where
$\gamma = 0.577215...$ is  the Euler-Mascheroni constant.
The non-zero values of $t^{(3)}_n$, obtained for $n \geq 6$, 
are given by a  more complicated expression, 
\begin{equation}
    t^{(3)}_n = \frac{ H_{n-2}^2 }{ 2n } - \frac{ H_{n-2} }{ n }
    + \frac{ 1 }{ n } \bigg ( -\frac{1}{3} + H_3 - \frac{1}{2} H_3^2 \bigg )
 + \frac{1}{n(n-2)} - \frac{1}{2n} \sum_{m=4}^{n-2} \frac{1}{m^2} ,
\end{equation}
with the first term on the r.h.s. determining the large $n$ asymptotics of 
$ t^{(3)}_n $. One can show that the large $n$ asymptotics of $ t^{(k)}_n $ 
is determined by the application of the recurrence relation 
(\ref{eq:t_k_n_recurrence}) to the leading term of $ t^{(k-1)}_n $. 
This is a consequence of the relation 
\begin{equation}\label{eq:sum_harmonic_number_j_over_j}
    \sum_{j=m}^n \frac{ H_j^p }{ j } 
    = \frac{ H_n^{p+1} - H_{m-1}^{p+1} }{ p+1 } 
    + O( H_n^{p-1} ) ,
\end{equation}
which is  analogue of the integral formula 
\begin{equation}
    \int_m^n \frac{ ( \ln x )^p }{ x } d x = \frac{ ( \ln n )^{p+1} - ( \ln m )^{p+1} }{ p+1 } .
\end{equation}
To prove Eq. (\ref{eq:sum_harmonic_number_j_over_j}) 
one can use the summation by parts formula of Eq.~(\ref{eq:summation_by_parts}) 
with  $u_j = H_j^p$, $w_j = j^{-1}$, and $W_j = H_j$. One obtains then  
\begin{equation}\begin{aligned}
    \sum_{j=m}^n \frac{ H_j^p }{ j } 
    & = \frac{ H_n^{p+1} - H_{m-1}^{p+1} }{ p+1 } \\
    & - \frac{ p }{ p+1 } \sum_{q=1}^p (-1)^q \binom{ p }{ q } \sum_{j=m}^n \frac{ H_j^{p-q} }{ j^{q+1} } \\
    & - \frac{ 1 }{ p+1 } \sum_{q=2}^p \binom{ p }{ q } \sum_{j=m-1}^{n-1} \frac{ H_j^{p-q+1} }{ (j+1)^q } . 
\end{aligned}\end{equation}
The sub-dominant contributions to this sum can be bounded by the inequalities
\begin{equation}
    \sum_{j=m}^n \frac{ H_j^p }{ j^q } \leq H_n^p \sum_{ j=m }^n \frac{ 1 }{ j^q } \leq \zeta(q) H_n^p , 
\end{equation}
where $q > 1$ and $\zeta(s)$ is the Riemann zeta function.
Employing the relation (\ref{eq:sum_harmonic_number_j_over_j}) 
while executing the recurrence  (\ref{eq:t_k_n_recurrence}) 
and replacing $H_n$ by $\ln n$ one obtains Eq.~(\ref{eq:tkn_asymptotics}).  
It is easy to see that the cumulative coefficients $d_N^{(K)}$ 
obtained by performing the summation of $t_n^{(k)}$ 
over  $k$ behave essentially in the same way  as  $t_n^{(k)}$
\begin{equation}\label{eq:d_K_N_asymptotics}
    d_N^{(K)} = \frac{ (\ln N)^{K-1} }{ N (K-1)! } \bigg[ + O \bigg( ( \ln N )^{-1} \bigg) \bigg] . 
\end{equation}
Note that the $N \rightarrow \infty$ limit (with $K$ fixed) of $d^{(K)}_N$ is zero. 
However, for large $K$ the values of $d^{(K)}_N$ are very close to $1/e$
for a quite broad range of $N$.  
For instance $1/e-d^{(20)}_N$=1.3$\times$$10^{-14}$,  7.3$\times$$10^{-8}$,  
 3.3$\times$$10^{-5}$, and 1.0$\times$$10^{-3}$ for $N=10^2,\, 10^3,\, 10^4$, 
and $10^5$, respectively. 

\end{document}